\documentclass[floatfix,preprint,aps,prd,showpacs,amsmath,amssymb]{revtex4}
\usepackage{graphicx}
\usepackage{float}
\usepackage{amsthm,amsfonts}
\begin{document}
\title{The influence of Unruh effect on quantum steering for accelerated two-level detectors with different measurements}

\author{Tonghua Liu$^{1}$, Jieci Wang$^{1,2}$\footnote{Email: jieciwang@iphy.ac.cn},  Jiliang Jing$^{1}$, and Heng Fan$^{2}$}
\affiliation{$^1$ Department of Physics, and Synergetic Innovation Center for Quantum Effects \\
and Applications,
 Hunan Normal University, Changsha, Hunan 410081, China\\
 $^2$  Institute
of Physics, Chinese Academy of Sciences, Beijing 100190, China
}

\begin{abstract}

\vspace*{0.2cm}  We study the dynamics of steering between two correlated  Unruh-Dewitt detectors when one of them locally interacts with external scalar field via different  quantifiers.  We find that the quantum steering, either measured by the entropic  steering inequality or the  Cavalcanti-Jones-Wiseman-Reid inequality,  is fragile under the influence of  Unruh thermal noise. The quantum steering is found always asymmetric and the asymmetry is extremely sensitive to the initial state parameter. In addition, the steering-type quantum correlations experience ``sudden death" for some accelerations, which are quite different from the behaviors of other quantum correlations in the same system. It is worth noting that the domination value of the tight quantum steering exists a transformation point with increasing acceleration. We also find that the robustness of quantum steerability  under  the Unruh thermal noise can be realized by  choosing  the smallest energy gap in the detectors.

 \end{abstract}

\pacs{04.62.+v, 03.70.+k, 06.20.-f, 05.30.Jp }

\maketitle
\section{Introduction.}

Quantum steering, an intermediate quantum correlation between  entanglement \cite{P.H.P}
and Bell nonlocality \cite{N.B.D},   was proposed by Schr\"{o}dinger in 1935 \cite{E.P.C, E.P.C1}.
Steering is a phenomenon that allows one party, Alice, to steer the state of a distant party Bob by exploiting the shared entanglement. This fascinating phenomenon  is regarded  at the heart of the Einstein-Podolksy-Rosen \cite{EPR} paradox and as a  characteristic trait of quantum mechanics.
Quantum steering has been intensively investigated in most recent years \cite{HMW,Handchen,SuN,steering7,Kocsis,Bowles,Adesso2015} after the realization that, apart from its  foundational importance in both fundamental and practical perspective of quantum information, it is advantageous to perform some  quantum information processing tasks  \cite{CBNG,HMW}. Despite many efforts on understanding the quantum steering \cite{dsteerable1,dsteerable2,dsteerable3, dsteerable4, dsteerable5,dissteer,dissteer1,dissteer2,dissteer4}, unlike entanglement and  nonlocality in which a variety of measures exist, there is still scarce literature concerning the quantification of the quantum steerability \cite{steeringq1,SUN,ANGELO}.

On the other hand, a famous achievement in quantum field theory is the Unruh effect \cite{unruh1976, unruhreview}, which indicates that quantum properties of fields are observer dependent.
The Unruh effect reveals the fact that an uniformly accelerated detector in the Minkowski vacuum will detect
thermal radiation at a temperature related to the proper acceleration of the detector.  While the mathematical  derivation of the Unruh effect is well established, direct observation of the Unruh effect in laboratories is still
lacking since an observable  Unruh temperature requires  currently experimental unreachable acceleration.  To get closer to  Unruh's  original derivation, people have modeled the observer by a  point like two-level atom, which is named  Unruh-DeWitt detector  \cite{UW84}. The detector is semiclassical because it possesses a classical world line while  its internal degree of freedom is treated quantum mechanically.  In addition, the study of  quantum information aspects of the Unruh effect is essentially related to  the information loss problem \cite{Bombelli-Callen,Hawking-Terashima} of black holes  because,  according to the equivalent principle, an accelerated observer in the  Minkowski vacuum corresponds to an observer who hovers outside the event horizon of a black hole \cite{wald94,RQI1,RQI2,RQI3,RQI4,RQI5}.

In this paper we propose  a  tight measure of steering to  analyze the quantum steering in a relativistically consistent way.  We also study different measures of quantum steering between a pair of Unruh-Dewitt detectors when one of them is accelerated. Here we employ the Unruh-Dewitt detector model \cite{UW84} rather than global free models \cite{SUN} to study the behavior of quantum correlations  in a relativistic setting  \cite{Landulfo, Landulfo1,tianwang,wangtian} because the latter suffers from the single-mode approximate problem and  physically unfeasible detection of quantum correlations in the full spacetime \cite{Bruschi}.  We find that the quantum steering, either measured by the entropic  steering inequality or the  Cavalcanti-Jones-Wiseman-Reid nonlocality  steering inequality,  is destroyed under the influence of the Unruh thermal noise.  In addition, the entropic  steering inequality measuring steering is found always  asymmetry for any acceleration.

The outline of the paper is as follows. In Sec. II, we introduce two types of measurements to quantify steering and define a tight measure of steering. In Sec. III, we discuss the quantum information description of the accelerated Unruh-Dewitt detector and the evolution of the prepared state under the Unruh thermal bath. In Sec. IV, we study the behaviors of quantum steering in the relativistic setting with different measurements. Conclusions are given in the last section.

\section{Quantifications of  quantum steering}


\subsection{ Quantum steering based on the entropic inequality}
Most recently,  a  quantification for quantum steering \cite{SUN} was performed based on the entropic uncertainty relations  \cite{dsteerable3}.
For a bipartite system $\rho_{AB}$ with two  observables  $\hat{R}$ and $\hat{O}$ in the $N$-dimensional Hilbert space, the quantum steering can be obtained by performing measurements on either the  subsystem $A$ ($A$-measuring) or the subsystem $B$ ($B$-measuring). We start with the entropic steering inequality for the $A$-measuring case, which  has the form \cite{dsteerable3}
\begin{equation}
H(R^B|R^A)+H(O^B|O^A)\geq -\log(\Omega^B),
\label{HAB}
\end{equation}
where $\Omega \equiv \max_{\{i,j\}}\{|\langle R_i|O_j\rangle|^2\}$, and $\Omega^B$ is the value $\Omega$ associated with observables $\hat{R}^B$ and $\hat{O}^B$.  $H(\cdot)$ is the von Neumann entropy, and $H(R^B|R^A)$ is the conditional entropy after  measurements on the subsystem A. In this paper we employ the measurements in the Pauli $X$, $Y$ and $Z$ bases on each side and consider a complete sets of pairwise complementary (mutually unbiased) observables. Then one  can obtain the following form of entropic steering inequality of $A\to B$
\begin{equation}
SI^{A\rightarrow B}=H(\sigma_x^B|\sigma_x^A)+H(\sigma_y^B|\sigma_y^A)+H(\sigma_z^B|\sigma_z^A)\geq 2.
\end{equation}
Basing on the entropic steering inequality, we  obtain the  steering of the $A$-measuring case to quantify how much the  bipartite state $\rho$ is steerable by Alice's measurements \cite{SUN}
\begin{equation}
S^{AB}:=\max\{0,\frac{{SI}^{A\rightarrow B}-2}{{SI}_{\max}-2}\},
\label{STAB}
\end{equation}
where $SI_{\max}$ is obtained  when the entropic steering inequality is maximal violated  for a given state and  equals to $6$ when the given state is a maximally entangled state \cite{SUN}. Similarly,  the  steering  for the $B$-measuring case can be defined by exchanging the roles of $A$ and $B$:
\begin{equation}
S^{BA}:=\max\{0,\frac{{SI}^{B\rightarrow A}-2}{{SI}_{\max}-2}\}.
\label{STBA}
\end{equation}

\subsection{Quantum steering based on the  CJWR  inequality }
Alternatively, Cavalcanti, Jones, Wiseman, and Reid (CJWR) developed \cite{dsteerable1} an inequality to judge whether a bipartite state is steerable. Later R. M. Angelo $et.\,al$
propose a measure of steering basing on the maximal violation of  the CJWR steering inequalities \cite{ANGELO}.
 The CJWR inequality for a bipartite state  $\rho$ has the form
\begin{equation}
F_n (\rho,\mu)=\frac{1}{\sqrt{n}} |\sum^n_{i=1}\langle A_i\otimes B_i\rangle|\leqslant 1,
\label{FCJ}
\end{equation}
where  $F_n$ is a real-valued function and $A_i=\vec{u}_i\cdot\vec{\sigma}$, $B_i=\vec{\upsilon}_i\cdot\vec{\sigma}$ are two sets of observables of the system. In Eq. (\ref{FCJ}), $\vec{\sigma}=({\sigma}_1,{\sigma}_2,{\sigma}_3)$ is the Pauli matrices, and $\vec{u}_i\in \mathbb{R}^3$ are unit vectors;  $\vec{\upsilon}_i\in \mathbb{R}^3$ are orthonormal vectors, and $\mu=\{\vec{u}_1,...,\vec{u}_n,\vec{\upsilon}_1,...,\vec{\upsilon}_n\}$ is a set of measurement directions.
Here we only consider the quantum system in which Alice and Rob are both allowed to measure two observables. More specifically, we assume that Alice and Rob each select anyone of two dichotomic  measurements  $\{A_1,A_2\},\{B_1,B_2\}$ for different directions and the outcomes of the observable $A$ are labeled $a\in \{-1,+1\}$ and similarly for other measurements.
For a two-observable quantum system, it has been proven in  \cite{ANGELO} that  the quantum steering can be quantified by
\begin{equation}
S_2:=\max \{0,\frac{F_2(\rho)-1}{F_2^{\max}-1}\},
\label{Sn}
\end{equation}
where $F_2(\rho)=\max_{\mu}[F_2(\rho,\mu)]$ is taken over all measurement settings $\mu$ and  $F^{\max}_2=\max_{\rho} [F_2 (\rho)]$ is the  maximal value  takes over all the bipartite states
\begin{equation}\label{CJWRq1}
F_2(\rho)=\sqrt{c^2-c^2_{\min}}.
\end{equation}
In Eq. (\ref{CJWRq1}) $c=\sqrt{\vec{c}^2}$,  $c_{\min}\equiv \min\{|c_1|,|c_2|,|c_3| \}$, and $c_i$ are components of the  bloch sphere vector $\vec{c}$ in bloch sphere   expansion  of  the state  $\rho$  \cite{LUO}
\begin{equation}
\rho = \frac{1}{4}(\mathbb{I}\otimes \mathbb{I}+\vec{a}\cdot \vec{\sigma}\otimes \mathbb{I}+\mathbb{I}\otimes \vec{b}\cdot\vec{\sigma}+\sum^3_{r=1}c_i\sigma_i\otimes\sigma_i),
\label{bloch}
\end{equation}
where $\mathbb{I}$ is the $2\times2$ unit matrix and $\{\vec{a},\vec{b},\vec{c}\}\in \mathbb{R}^3$ are vectors with norm length smaller than $\mathbb{I}$.
The value of  $F_2^{\max}$  can be  determined by using the inequality $\vec{a}^2+\vec{b}^2+\vec{c}^2\leqslant 3$  with $\vec{a}^2=\vec{b}^2=0$. Hence, $F_2^{\max}=\sqrt{2}$  \cite{dsteerable1} .

\subsection{Tight measure of steering}

In the last two subsections we have introduced two types of measurements for quantum steering based on the  entropic steering inequality and  the CJWR  inequality, respectively. However,  it is worth noting that there may be  different types of steering  inequalities for a given quantum state and each type of them is not  superior to all other forms of steering inequalities. For example,  for some states the steering can be  measured by  the violation of the entropic steering inequality but can't be measured by  the  violation of  the CJWR inequalities. In other words, the state may violate  entropic steering inequality but  fail to violate CJWR steering inequality, vice versa. In addition, the first measure fails  to give  united amount of quantum steerability for different measuring directions, which makes the quantum steering inherent  asymmetric. On the other hand, the  second measures can't distinguish  the notions of steering and Bell nonlocality  in the two-measurement scenario \cite{ANGELO}.
To obtain an united quantifier,  we propose  a more tight limit for the measure of steering based on the aforementioned quantifiers. That is, for a given state, it is steerable only if all the above mentioned steering inequalities are violated with the same set of measurements. Therefore, the tight quantum steerability of a bipartite system is defined by minimuming  the   violation over all the  inequalities, which is
\begin{equation}\label{gsteering}
 \textsf{S}_g\equiv \min\{S^{1}, S^{2},... S^n\}.
\end{equation}
where $S^{1}, S^{2},... S^n$ are different quantifiers basing on the violation of different steering inequalities for the same state. Here we have  $S^{1}, S^{2},.... S^n=S^{AB}, S^{BA}, S_2$.

\section{Evolution of the detectors' state  under  relativistic motion}

In this section we give a brief description for  the Unruh-Dewitt detectors \cite{UW84} from a quantum information viewpoint and discuss the evolution of the system when one detector experiences relativistic motion. The Unruh-Dewitt detector is modeled by a  two-level atom which interacts only with its nearby fields. We assume that the  detectors are initially share some quantum correlations  in the Minkowski spacetime and are observed by Alice and Bob, respectively. The atom carried by Alice keeps inertial and is always  switched off while  Rob's detector interacts with the scalar field when it  moves with uniform acceleration  for a time duration $\Delta$.  The world line of Rob's detector is
\begin{equation}\label{worldline}
t(\tau)=a^{-1}\sinh a\tau,\;x(\tau)=a^{-1}\cosh a\tau,
\end{equation}
$y(\tau)=z(\tau)=0$, where $a$ is detector's proper acceleration. In Eq. (\ref{worldline})  $\tau$ is  proper time of the detector and $(t,x,y,z)$ are the usual Cartesian coordinates in the Minkowski spacetime. Throughout this paper, we employ natural units $c=\hbar=\kappa_{B}=1$. Since Rob's detector interacts with the field, the interaction should be taken into account  and  the total system evolves to a tripartite one. The initial
state of the total system is
\begin{equation}
|\Psi_{t_0}^{AR\phi}\rangle=|\Psi_{AR}\rangle\otimes|0_{M}\rangle
,\label{IS}%
\end{equation}
where $|\Psi_{AR}\rangle=\sin\theta |0_{A}\rangle|1_{R}\rangle+\cos\theta|1_{A}\rangle
|0_{R}\rangle$ denotes the initial state shared  by Alice's (A) and Rob's (R) detectors, and $|0_{M}\rangle$ represent that the external scalar field is in Minkowski vacuum.

The  total Hamiltonian of the system is given by
\begin{equation}
H_{A\, R\, \phi} = H_A + H_R + H_{KG} +  H^{R\phi}_{\rm int},\label{totalh}
\end{equation}
where $H_{A}=\Omega A^{\dagger}A$  and $H_{R}=\Omega R^{\dagger}R$ are  Hamiltonians of the detectors and   $H_{KG}$ is   Hamiltonian of the external scalar field; $\Omega$ is the energy gap of the detectors. In Eq. (\ref{totalh}) the
interaction Hamiltonian $H^{R\phi}_{\rm int}(t)$, which describes how Rob's detector is coupled with the external massless scalar field $\phi(x)$, is given by
\begin{equation}
H^{R\phi}_{\rm int}(t)=
\epsilon(t) \int_{\Sigma_t} d^3 {\bf x} \sqrt{-g} \phi(x) [\chi({\bf x})R +
                           \overline{\chi}({\bf x})R^{\dagger}],
\label{int}
\end{equation}
where $g\equiv {\rm det} (g_{ab})$,  $g_{ab}$ is the Minkowski  metric, and $\bf{x}$ are coordinates defined on the Cauchy surface $\sum$ associated with the timelike isometries followed by the qubits.  In Eq. (\ref{int}), $\epsilon(t)$ is a real-valued switching function which keeps the detector  switched on smoothly for a finite amount of proper time $\Delta$ and $\chi(\mathbf{x})=(\kappa\sqrt{2\pi})^{-3}\exp(-\mathbf{x}^{2}/2\kappa^{2})$  is a point-like coupling function which  guarantees  the detector is  space-localized and only interacts with the field in a neighborhood of its world line.
By combining the detector-field free Hamiltonian as $H_0 = H_A + H_R + H_{KG}$, one can cast the total Hamiltonian as
\begin{equation}
H_{R\, \phi}= H_0 + H_{\rm int}.
\end{equation}
The  state $|\Psi^{AR \phi}_t \rangle$  at time $t=t_0+\Delta$ can be written as
\begin{equation}
|\Psi^{AR \phi}_t \rangle =
T \exp[-i\int_{t_0}^t dt H_{\rm int} ^I(t)] |\Psi^{AR \phi}_{t_0} \rangle,
\label{Dyson1}
\end{equation}
in the interaction picture, where $T$ is the time-ordering operator.  In Eq. (\ref{Dyson1})$,
H_{\rm int}^I (t) = U^{\dagger}_0(t) H_{\rm int} (t) U_0 (t)$  and
 $U_0 (t)$ is the unitary evolution operator associated with the  Hamiltonian
$H_0 (t)$. Then the final state
$|\Psi^{AR \phi}_{t} \rangle $ is found to be
\begin{widetext}
\begin{eqnarray*}
|\Psi^{AR \phi}_{t} \rangle
 &=& T\exp [-i\epsilon(t)\int d^4x\sqrt{-g}U^{\dagger}_0(t)\phi(x)(\chi({\bf x})R+\overline{\chi}({\bf x})R^{\dagger })U_0(t)]|\Psi^{AR \phi}_{t_0} \rangle \\
 &=& T \exp [-i\epsilon(t)\int d^4x\sqrt{-g}U^{\dagger}_0(t)\phi(x)U_0(t)(\chi({\bf x})e^{-i\Omega t}e^{i\Omega t}R+\overline{\chi}({\bf x})e^{i\Omega t}e^{-i\Omega t}R^{\dagger })]|\Psi^{AR \phi}_{t_0} \rangle
\\
 &=& T \exp [-i\int d^4x\sqrt{-g}\phi^I(x) (fR^I + \overline{f}R^{\dagger I})] |\Psi^{AR \phi}_{t_0} \rangle,
  \label{Dyson2}
\end{eqnarray*}
\end{widetext}
where
$f \equiv \epsilon(t) e^{-i\Omega t}\chi ({\bf x})$ and $R^I=e^{i\Omega t}R$ have been used. The dynamics of the atom-field system at time $t=t_0+\Delta$ can be calculated  in the first order of perturbation over the coupling constant $\epsilon$ \cite{UW84}. Under the dynamic evolution described by the Hamiltonian given by Eq.(\ref{int}), the final state of the system
  is
 \cite{Landulfo,Landulfo1,tianwang,wangtian,wald94}
\begin{equation}
|\Psi^{A R \phi}_{t} \rangle
= [I - i(\phi(f)R + \phi(f)^{\dagger} R^{\dagger}) ] |\Psi^{AR \phi}_{t_0} \rangle,
\label{primeira_ordem}
\end{equation}
where the  operator
\begin{eqnarray}
\nonumber\phi(f) &\equiv& \int d^4 x \sqrt{-g}\chi(x)f\\&=&i [a_{RI}(\overline{u E\overline{f}})-a_{RI}^{\dagger}(u Ef)],
\label{phi(f)}
\end{eqnarray}
is the distribution  function of the  external scalar field.
In Eq. (\ref{phi(f)}), $Ef$ is approximately a positive-frequency solution of the scalar field \cite{Landulfo,Landulfo1,tianwang,wangtian,wald94}.
Besides, the $u$ operator is the positive-frequency part of the solutions of K-G equation with respect to the timelike isometry, and
\begin{equation}
Ef = \int d^4x'\sqrt{-g(x')} [G^{\rm adv}(x, x')-G^{\rm ret}(x, x')] f(x'),
\label{Ef}
\end{equation}
where $G^{\rm adv}$ and $ G^{\rm ret}$ are the advanced and retarded Green functions. The annihilation operator $a_{RI}(\overline{u})$ annihilates the Rindler vacuum for the single  mode  $u$, which is an positive-frequency solution of the Klein-Gordon equation in  the Rindler metric  \cite{Landulfo,Landulfo1,tianwang,wangtian,wald94}.

By  composing the  initial state Eq. (\ref{IS}) and Eq. (\ref{primeira_ordem}),
 the final state of the total system can be expressed  in terms of the Rindler operators $a_{R I}^{\dagger}$ and $a_{R I}$
\begin{eqnarray}
| \Psi^{AR \phi}_{t}\rangle
& = &
|\Psi^{AR \phi}_{t_0} \rangle
 + \sin \theta |0_A\rangle  |0_R\rangle
 \otimes(a_{R I}^{\dagger}(\lambda)|0_M\rangle)
 \nonumber \\
& + & \cos \theta |1_A\rangle |1_R\rangle\otimes(a_{R I}(\overline{\lambda})|0_M\rangle),
\label{evolutionAUX}
\end{eqnarray}
where $\lambda = -uEf$. In Eq. (\ref{evolutionAUX} ), the creation and annihilation
operators $a_{R I}^{\dagger}(\lambda)$ and $a_{R I}(\overline{\lambda)}$  are defined in the Rindler region $I$, while   $|0_M\rangle$ is the   Minkowski vacuum.  The relations between these two sets of
operators are \cite{Landulfo, wangtian}
\begin{eqnarray}
a_{R I}(\overline{\lambda})&=&
\frac{a_M(\overline{F_{1 \Omega}})+
e^{-\pi \Omega/a} a_M ^{\dagger} (F_{2 \Omega})}{(1- e^{-2\pi\Omega/a})^{{1}/{2}}},
\label{aniq} \\
a^{\dagger}_{R I}(\lambda)&=&
\frac{a^{\dagger}_M (F_{1 \Omega}) +
e^{-\pi \Omega/a}a_M(\overline{F_{2 \Omega}})}{(1- e^{-2\pi\Omega/a})^{{1}/{2}}}
\label{cria},
\end{eqnarray}
where
$F_{1 \Omega}=
\frac{\lambda+ e^{-\pi\Omega/a}\lambda\circ w}{(1- e^{-2\pi\Omega/a})^{{1}/{2}}}$, and
$F_{2 \Omega}=
\frac{\overline{\lambda\circ w}+ e^{-\pi\Omega/a}\overline{\lambda}}{(1- e^{-2\pi\Omega/a})^{{1}/{2}}}$. Here
$w(t, x)=(-t, -x)$
is a wedge reflection isometry that reflects  $\lambda$ defined in  the Rindler region $I$ into $\lambda\circ w$ in the other region $II$
\cite{Landulfo, wangtian,wald94}.

We are interested in the evolution of the detectors' state after interacting with the field. Therefore, the  part for the external field $\phi(f)$ should be  traced  out. Then we obtain
\begin{eqnarray}
\rho_{t}^{AR} = \left(
                  \begin{array}{cccc}
                    \gamma & 0 & 0 & 0 \\
                    0 & 2\alpha \sin^2 \theta  & \alpha\sin 2\theta  & 0 \\
                    0 & \alpha\sin 2\theta  & 2\alpha\cos^2 \theta  & 0 \\
                    0 & 0 & 0 & \beta \\
                  \end{array}
                \right)
,
\label{rhof2}
\end{eqnarray}
for the detectors, where  the parameters
$\alpha$, $\beta$ and $\gamma$ have the following forms
\begin{eqnarray}\label{pabc}
\nonumber \alpha  &  =\frac{1-q}{2(1-q)+2\nu^{2}(\sin^2 \theta+q\cos^2\theta)},\\
\nonumber \beta  &  =\frac{\nu^{2}q\cos^2\theta}{(1-q)+\nu^{2}(\sin^2 \theta+q\cos^2 \theta)},\\
\gamma  &  =\frac{\nu^{2}\sin^2\theta}{(1-q)+\nu^{2}(\sin^2 \theta+q\cos^2 \theta)},
\end{eqnarray}
respectively. In Eq. (\ref{rhof2}) the acceleration $a$ has been parametrized as
$q\equiv e^{-2\pi\Omega/a}$ and $\nu$ is  a combined coupling parameter, which is
$\nu^{2}\equiv||\lambda||^{2}=\frac{\epsilon^{2}\Omega\Delta}{2\pi}%
e^{-\Omega^{2}\kappa^{2}}$  \cite{Landulfo,Landulfo1,tianwang,wangtian}. In our model  $\epsilon\ll1$ and  $\nu^2\ll1$ are required  for the  validity of the perturbation approach.

\section{BEHAVIORS OF QUANTUM STEERING under the influence of Unruh thermal noise }

\begin{figure*}[tbp]
\centerline{\includegraphics[width=16cm]{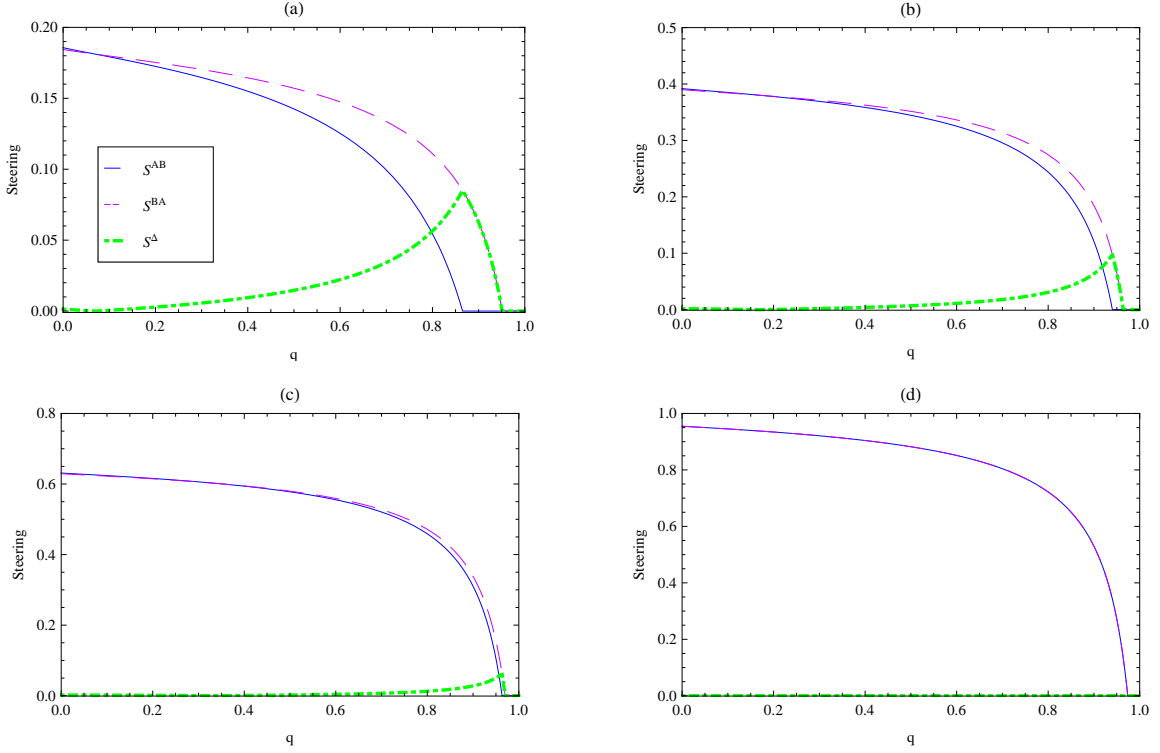}}
\caption{Steering of $S^{A B}$(blue solid line), $S^{B A}$(purple dash line), steering asymmetry $S^{\Delta}$(green dash line  between the detectors as a function of acceleration parameter $q$. The initial state parameter $\theta$ and effective coupling parameter $\nu$ are:  (a) $ \nu=0.1, \theta=\pi/12$, (b) $ \nu=0.1, \theta=\pi/8$, (c) $ \nu=0.1, \theta=\pi/6$,  (d) $ \nu=0.1, \theta=\pi/4$, respectively.}
\label{sb1}
\end{figure*}

In this section we study the behaviors of quantum steering under the influence of thermal field induced by the Unruh radiation.  The initial state of the entire  system is given in Eq. (\ref{IS}), from which we can see that the amount of the detectors'  initial  state $|\Psi_{AR}\rangle$  depends on $\theta$. For any nonzero $\theta$, the detector $A$ shares initial quantum correlation with detector $R$. Then Rob's  detector is accelerated for a time duration $\Delta$ with constant acceleration and influenced by the Unruh thermal bath. We are interested in if the steering-type quantum correlations between the detectors is destroyed by the Unruh thermal noise and if the asymmetric of quantum steering from different directions is affected by the Unruh effect.

\subsection{Behaviors of quantum steering measured by the violation of the entropic inequality}

We start with discussing the dynamics of  quantum steerability measured by  the violation of the entropic  steering inequality under the influence of the Unruh radiation.
We aware that the dynamics of entropic inequality  measuring  steering has been discussed for free bosonic field modes \cite{SUN}. However,  the quantum correlations for free    bosonic modes  suffers from two disadvantages in a relativistic frame.  The first one is the so called single-mode approximate  problem  \cite{Bruschi}, that is, one cannot map a single-frequency  Minkowski mode  with a single-frequency Rindler mode. This is because an inertial observer in the Minkowski spacetime is freely to create excitations in any accessible
modes in the accelerated frame \cite{Bruschi}.   Therefore, we have to choose the  Minkowski modes as superpositions of
different frequencies of Rindler modes.  Secondly, the free field quantum correlation  suffers from the unfeasible detection in the full spacetime because the spacetime is isolated by the Rindler event horizon. To overcome these disadvantages, here we employ the  Unruh-Dewitt detector which  interacts only with its nearby fields and locally detectable.

Since the  violation of the entropic  uncertainty relations measuring quantum steering depends on the direction of the performed measurements, it may be asymmetric \cite{Adesso2015} under different measuring parts. To better understand this property, we calculate the entropic steering $S^{A B}$ for the  $A\to B$ direction  and  $S^{BA}$ for the $B\to A$ direction for the final state. To check the degree of steerability asymmetric, we also calculate the steering asymmetry, which is  $S^{\Delta}=|S^{AB}-S^{B A}|$.

In Fig. (1) we plot the steering of $S^{A B}$, $S^{BA}$, as well as the steering asymmetry $S^{\Delta}$ versus the acceleration parameter $q$ for fixed coupling parameter  $\nu=0.1$ and different initial state parameters. It is shown that both the quantum steering of $S^{A B}$ and $S^{BA}$  monotone degrade with the growth of the acceleration parameter $q$, which means that the thermal noise induced by Unruh radiation destroys  the steering-type quantum resources.
It is worth noting  that the steering-type quantum correlations happen ``sudden death" for some acceleration $q$, which is quite different form the behavior of discord-type correlation where the quantum discord has a ``sudden change" point and approaches zero only in the limit of $q\to 1$  \cite{Landulfo1}.
It is  shown that the $B\to A$ steering  is always bigger than the $A\to B$  with growing acceleration except the case of $\theta=\pi/4$. Considering that they  equal to each other when there are no acceleration ($q=0$),  we can conclude that the steering form the noninertial part to  the inertial part is less affected by the Unruh thermal noise.

From Figs. (1a-1d) we can see that the steering asymmetry increases   and  then decreases with increasing acceleration parameter $q$. In addition,  the maximal steering asymmetry appears when one of the two steering directions suffers from ``sudden death".  Furthermore, the  ``sudden death" point is in fact a  critical point  because only one direction is steerable after the one-way  ``sudden death".  In other words, the system is currently experiencing a transformation under the influence of Unruh effect and  the transformation will happen earlier for a smaller initial state parameter $\theta$.
This result is nontrivial because it is shown that the asymmetry of steerability is extremely sensitive to the initial state parameter, which is a unique nature of quantum steering comparing with  the behaviors of other forms of quantum correlations, such as quantum entanglement \cite{Landulfo}, quantum discord \cite{Landulfo1},   quantum nonlocality \cite{tianwang} and quantum coherence \cite{wangtian} in the same system.  We can see from Fig. (1d) that the steering of $S^{AB}$ is almost equivalent to $S^{BA}$, which means that the asymmetry   of steering is very small   when the initial state is a maximally entangled state. 

We note that the effect of thermal noise on the quantum steering between two systems has previously been studied in  \cite{QYH,QYHS,QYHL}.  For example, the authors examined the effect of an initial thermal excitation of an oscillator on observing an EPR paradox between an oscillator and a pulse in Ref.  \cite{QYH}. They found that the steering suffers ``sudden death"  for above a certain threshold value for thermal noise. Here we find that, similar with results for Gaussian states \cite{QYH,QYHS}, the thermal noise induced by the Unruh effect  damages the steering and the ``sudden death" also found for steering-type correlation. However, the source of thermal noise is attribute to thermal
occupation of the mechanical oscillator, but the generation of thermal noise is come from
the atom¡¯s acceleration of the detectors in our present paper. We also find that in the relativistic case the maximal steering asymmetry
appears when one of the two steering directions suffers from``sudden death". In addition,  the thermal bath induced by the Unruh radiation always generates asymmetry of steerability. These behaviors are not found in  earlier papers in inertial systems.

\subsection{Behaviors of CJWR  inequality measure of quantum steering and the tight measure of steering }

Now we have discussed the dynamics of  entropic inequality measuring quantum steering between an inertial detector and an accelerated Unruh-Dewitt detector. However, as we stated before,  this measure of steering fails  to give  united amount of quantum steerability for different measuring directions.  In this subsection we  calculate the CJWR  inequality measure of quantum steering and the tight measure of steering for the final state Eq. (\ref{rhof2}) after the accelerated motion of Rob's detector.
To this end we rewrite the final state
Eq. (\ref{rhof2})  to a  Bloch sphere  expansion  form with vectors  $\vec{a}=(a_1,a_2,a_3),\vec{b}=(b_1,b_2,b_3),\vec{c}=(c_1,c_2,c_3)$.
After some calculations the final state is given by a   general form of bloch sphere  expansion
\begin{widetext}
\begin{eqnarray}
\rho_{t}^{AR}= \frac{1}{4}\left(
                  \begin{array}{cccc}
                    a_3+b_3+c_3+1 & b_1-ib_2 & a_1-ia_2 & c_1-c_2 \\
                    b_1+ib_2 & a_3-b_3-c_3+1  & c_1+c_2  & a_1-ia_2 \\
                    a_1+ia_2 & c_1+c_2  & -a_3+b_3-c_3+1  & b_1+ib_2 \\
                    c_1-c_2 & a_1+ia_2 & b_1+ib_2 & -a_3-b_3+c_3+1 \\
                  \end{array}
                \right),
\end{eqnarray}
\end{widetext}
where
\begin{eqnarray}
\nonumber a_1&=&a_2=b_1=b_2=0,\\
\nonumber c_1&=&c_2=2\alpha\sin{2\theta},\\
    a_3&=&4\alpha\sin^2{\theta}+2\gamma-1,\\
\nonumber b_3&=&4\alpha\cos^2{\theta}+2\gamma-1,\\
 \nonumber c_3&=&1-4\alpha.
\end{eqnarray}
Using Eqs. (\ref{Sn})  and (\ref{CJWRq1}) we  obtain the  CJWR  inequality measure of quantum steering  $S_{2}$ and plot them in Fig. (2).

\begin{figure}[ht]
\centerline{\includegraphics[scale=0.65]{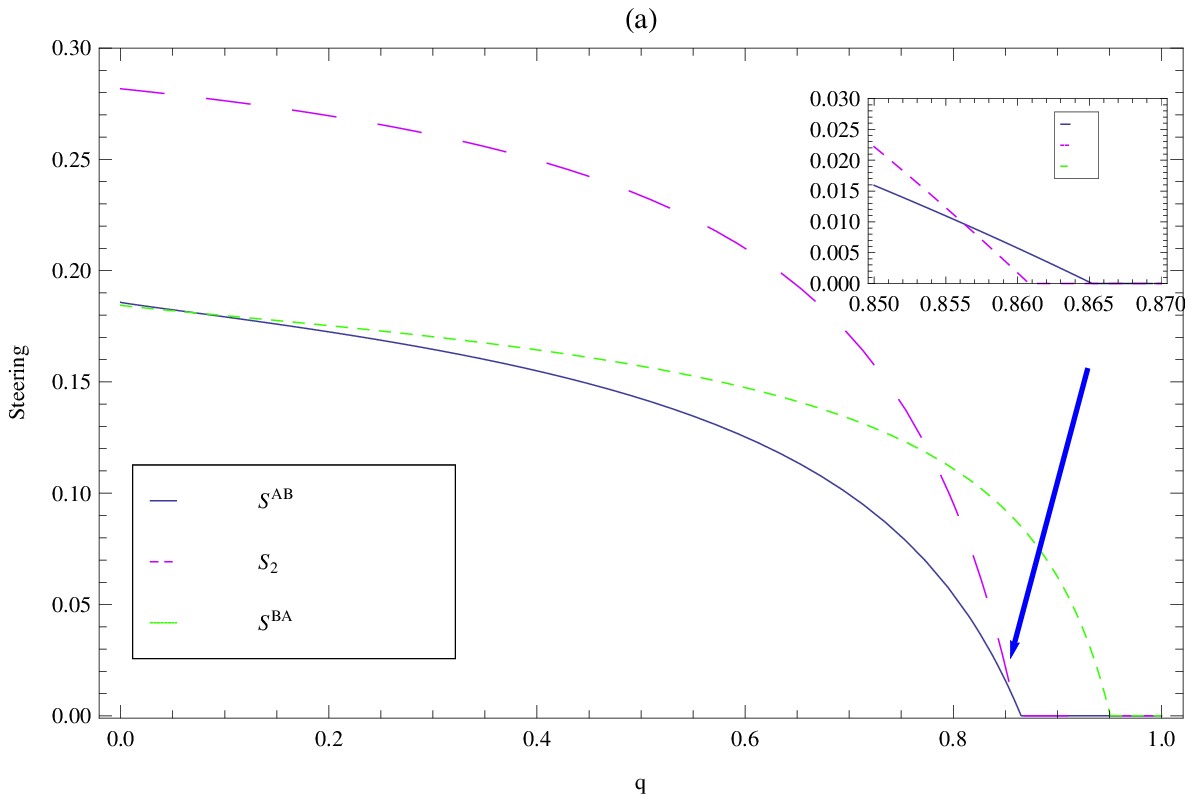}}
\includegraphics[scale=0.54]{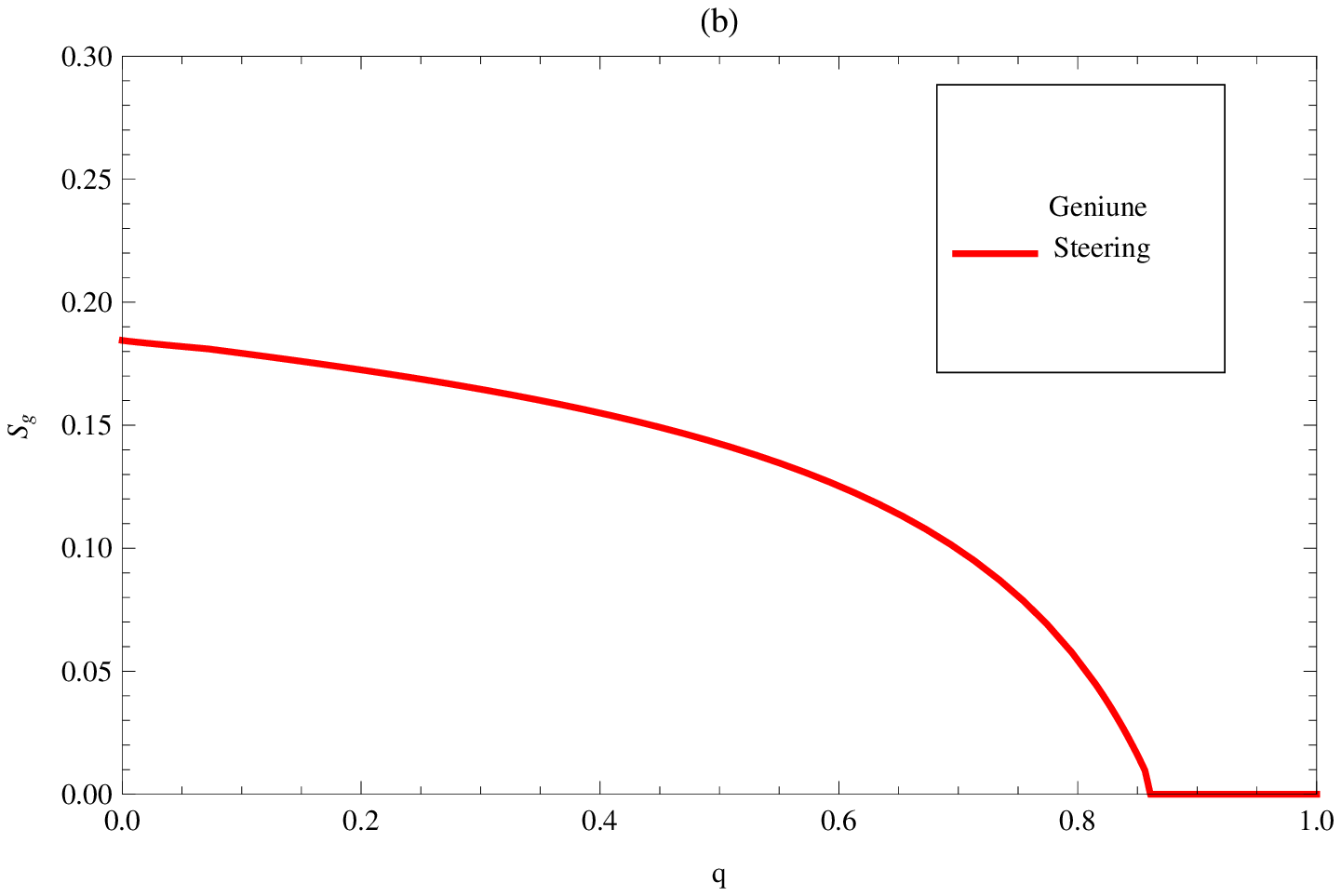}
\caption{\label{ClaT}(a) The quantum steering of  $S_{2}$, $S^{A B}$, and $S^{B A}$ as a function of the acceleration parameter $q$. (b) The tight steering $\textsf{S}_{g}$  between Alice and Rob as a function of the detector acceleration parameter $q$. The initial state parameter $\theta=\pi/12$ and effective coupling parameter $\nu=0.1$ }
\end{figure}

In Fig. (2a) we plot the behavior of the $S_{2}$ steering and compare it with the   $S^{A B}$ and $S^{B A}$ steering. All the three measurements are plotted as a function of the acceleration parameter $q$ for fixed   coupling parameter $\nu=0.1$ and initial state parameter $\theta=\pi/12$.
The  variation trend of the curve of steering $S_2$ is similar with the entropic steering  $S^{AB}$ and $S^{B A}$ and it also experiences ``sudden death" for some  accelerations.  We find that both of the measures of steering  monotone decrease with the increase of the acceleration parameter $q$. Here we have employed two types of quantifier for the steering of the final state and obtained similar results.  It is also found that the  CJWR inequality measure of quantum steering can only  measure the degree of a bipartite steerable state but can't specific show the asysmmetic of quantum steering. This  is because the steering inequality of the CJWR  measure  is equivalent with the inequality of Bell nonlocality  where the system allows to measure two observables, which makes the notions of steering and Bell nonlocality  derived  here  indistinguishable in the two-measurement scenario.

Now let us give some physical interpretations for the loss of quantum  steering in the accelerated system.    Alice performs trusted quantum measurements on her own subsystem, and Rob trusts his measurement apparatus.   At this moment Alice is able to convince Rob (who does not trust Alice) because they share more  quantum correlations initially.  However,  Alice gradually loses this ability when Rob's acceleration is growing because the shared quantum correlations are decreased under the Unruh thermal noise.

It is worth noting that the  steering $S_2$ is more than the steering $S^{A B}$ and $S^{BA}$ for some small accelerations, but it deceases much more  rapid than the latter two and appears an earlier ``sudden death'' with increasing acceleration. That is to say, there is no domination relation between these two types of steering quantifiers. Considering that the  tight measure of steering $\textsf{S}_{g}$  is defined by minimuming  the   violation over all the  inequalities, as has been shown in Eq. (\ref{gsteering}),  $\textsf{S}_{g}$ should take different parts from all the three steering measurements  $S_2$,  $S^{A B}$ and $S^{BA}$ for different accelerations.

In Fig. (2-b), we plot the tight measure of steering $\textsf{S}_{g}$ under the Unruh thermal noise as a function of the acceleration parameter $q$ for fixed coupling parameter $\nu=0.1$. We can see that the value of the tight quantum steering is dominated by the  $S^{A B}$ steering for small acceleration, but the  $S_2$ steering predominates the tight quantum steering when the acceleration parameter $q > 0.857$. That is to say, $q = 0.857$ is a critical point for the measure of tight quantum steering. We find again the tight quantum steering decreases with increasing acceleration of Rob's detector and appears ``sudden death'' . Then we arrive at the conclusion that the steering-type of quantum correlation is destroyed by the thermal bath induced by the Unruh effect.

\begin{figure}[ht]
\centerline{
\includegraphics[scale=0.7]{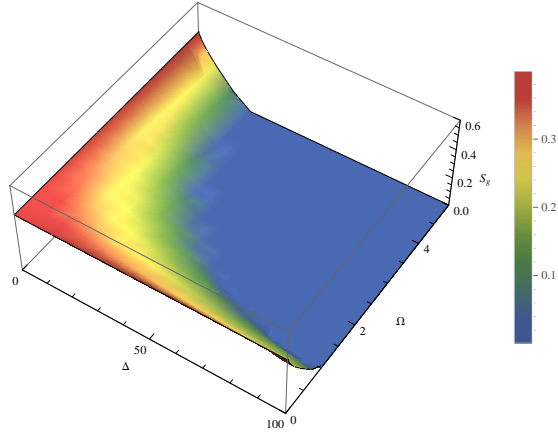}}
\caption{\label{ClaT1} The tight steering $\textsf{S}_{g}$  between Alice and Rob as  functions of the interaction time duration  $\Delta$
 and  the energy gap $\Omega$. The parameters related to the effective coupling parameter $\nu$ are  fixed as  $\epsilon=2\pi\cdot 10^{-4}$ and $\kappa=0.02$, respectively. The initial state parameter is $\theta=\pi/8$ and the acceleration parameter is  $q=0.5$. }
\end{figure}

We are also interested in how the detail of the interaction between the accelerated detector and external scalar field influences the steering-type quantum correlation. To this end we plot the tight steering $\textsf{S}_{g}$  of the system as functions of the energy gap $\Omega$ of Rob's detector and the interaction time $\Delta$ in Fig. (3). For the  validity of the perturbation approach applied in this work, the parameters related to the effective coupling parameter $\nu$ are fixed to satisfy $\epsilon\ll\Omega^{-1}\ll\Delta$. It is  shown in Fig. (3) that the steering-type quantum correlation is sensitive to the variation of different energy gaps  of the accelerated detector. In particular, the quantum steering is much more robust over the interaction time  when the energy gaps are small. On the other hand, the tight steering is very fragile under the interaction between the detector and the external field.  The energy gap of the accelerated two-level atom has  significant impact on the available quantum resource of the system. Therefore, we can prepare  proper detectors via some artificial two-level atoms that possess  proper  energy gaps to obtain robust steering-type quantum correlation over the Unruh thermal noise. Alternatively, larger values of steering are obtained for a shorter interaction time.

\vspace*{0.5cm}
\section{Conclusions}
In conclusion,  we have studied the evolution of  steering  for two entangled Unruh-Dewitt detectors when one of them is accelerated and interacts with  the neighbor external scalar field.
We employ two different measures of steering  based  on the violation of the entropic  steering inequality and the CJWR steering inequality, respectively.  Then we  define the tight steering
basing on the minimal violation over all the  inequalities. We find that the Unruh thermal noise will  destroy the steering-type quantum resource for all the  steering measurements. For the entropic inequality measure of quantum steering, it  exists asymmetric property and the asymmetry of steerability is extremely sensitive to the initial state parameter $\theta$. It worth noting that the steering-type quantum correlations happen ``sudden death'', which is quite different form the behavior of the discord-type correlation  in the limit of $q\to 1$. We find that the CJWR inequality measuring steering fails to specific how the asysmmetic of quantum steering and is  indistinguishable from the  Bell
nonlocality. In addition, the tight measure of steering $\textsf{S}_{g}$  should take different parts from all the three steering measurements  $S_2$,  $S^{A B}$ and $S^{BA}$ for different accelerations. The domination value of the tight quantum steering is found to have a transformation point $q =0.857$ for the measurements of steering-type quantum correlation. It is also shown that robust  tight steering  under  the  Unruh thermal noise can be obtained by  choosing the shortest interaction time allowed in quantum mechanics and  some small energy gaps.
We know that an accelerated observer in the Minkowski vacuum corresponds to a static observers outside a black hole in the Hartle-Hawking vacuum \cite{UW84, RQI1,Landulfo1}.  Similarly, a static observer in the Minkowski space-time corresponds to  a free-falling observer  in the Schwarzschild spacetime. Therefore, the analysis used to derive the results of our manuscript can, in principle, be applied to study  the dynamic behavior of quantum steering under the influence of Hawking radiation.

\begin{acknowledgments}
This work is supported by the National Natural Science Foundation
of China under Grant  No. 11675052, No. 11475061, and No. 91536108, the Doctoral Scientific Fund Project of the Ministry of Education of China under Grant No. 20134306120003, and the Postdoctoral Science Foundation of China under Grant No. 2014M560129, No. 2015T80146.

\end{acknowledgments}

\end{document}